\documentclass{epl2}

\usepackage{epsfig}

\title{Short-time quantum correlations in the atom
optics kicked rotor}

\author{S.A. Wayper, W. Simpson and M.D. Hoogerland}
\institute{Department of Physics, University of Auckland, Private Bag 92019,
Auckland, New Zealand}
\shortauthor{S.A. Wayper {\em et. al.}}


\abstract{
We experimentally verify the analytical expressions that exist for
the diffusion rate in the quantum delta kicked rotor system for small numbers
of kicks. We show development of diffusion resonances from two to 
five kicks, and of multiple resonances for high kick strengths. Furthermore,
we show that, in contrast to classical predictions, the results are
purely periodic in the kick period, and reproduce the predicted  quantum- and
diffusion resonances.}

\begin{document}
\maketitle
\newcommand{\kbar}{\mathchar'26\mkern-9muk}

The atom optics implementation of the delta kicked rotor system has
been the vehicle of a number of `Quantum Chaos' experiments:
experiments that study a quantum system whose classical equivalent
exhibits chaotic behavior. One such system is the classical
delta
kicked rotor system. The atom optics implementation of this
system
consists of a two-level atom placed in a pulsed standing wave of laser
light, which is detuned from resonance~\cite{Raizen1995}. The AC Stark shift
induces a
potential which depends sinusoidally on the position in the standing
wave. The atoms generally have a momentum distribution that is much
wider than the recoil of a single photon, and are distributed over a
large range of positions in the standing wave. The chaotic nature of
the classical equivalent of this system means that the energy growth
would be constant for each kick. The quantum mechanical nature of the
atomic motion gives rise to coherences, which curb the energy growth.

The Hamiltonian for the atomic motion in the pulsed standing wave
can be written as \cite{GrahamZoller}
\begin{equation}
\hat H=\frac{\hat p^2}{2m}-\frac{\hbar\Omega_R}{2}\cos(2k_l\hat
x)\sum_{n=1}^N f(t-nT)
\end{equation}
where $\hat p$ and $\hat x$ are the atomic momentum and position
operators, $\Omega_R=\Omega^2/(4\Delta)$ is the effective Rabi frequency,
$k_l$ is the
laser wave number, $m$ is the atomic mass
and $T$ is the kick period. The function $f(t)$
describes the laser pulse, and is generally given by a square pulse
with a length $\tau_p\ll T$.

The Hamiltonian can be scaled to dimensionless parameters to read
\begin{equation}
\hat H'=\frac{\hat\rho^2}{2}+\kappa\cos(\hat\phi)\sum_{n=1}^{N}f(\tau-n)
\end{equation}
where the scaled momentum operator $\hat\rho=2k_lT\hat p/m$, the scaled
position operator $\hat\phi=2k_l\hat x$, $\tau=t/T$ and  $\hat H'= (4
k_l^2T^2/m)\hat H$. The commutator $[\hat\phi,\hat\rho]=8i\omega_r
T\equiv i\kbar$, where $\omega_r=\hbar k_l^2/(2m)$ is the recoil
frequency. Consequently $\kbar$, which can be experimentally altered
by changing the kick period, plays the role of Planck's constant in the
uncertainty principle. Here, have also introduced the classical kick strength 
$\kappa=4\Omega_R\omega_r\tau_pT$.

This allows us to map out the transition from
`classical' ($\kbar \rightarrow 0$ at constant $\kappa$) to `quantum' ($\kbar \gg 1$) using a
convenient experimental parameter, the kick period. For convenience, we will also
introduce the experimental kick strength $\phi_d=\kappa/\kbar=\Omega_R\tau_p/2$.

Analytical results~\cite{analytic} exist for the values of the energy
after small numbers of kicks. For the energies after the first five
kicks we obtain:
\begin{eqnarray}
E_1 & = & E_0 + \phi_d^2 \nonumber \\
E_2 & = & E_0 + 2\phi_d^2 \nonumber \\
E_3 & = & E_0 + \phi_d^2\left(3-2J_2(\kappa_q)\right) \nonumber \\
E_4 & = & E_0 + \phi_d^2\left(4-4J_2(\kappa_q) +
2J_3(\kappa_q)-2J_1^2(\kappa_q)\right) \nonumber \\
E_5 & \approx & E_0 + 
\phi_d^2\left(5-6J_2(\kappa_q)+4J_3^2(\kappa_q)\right.\nonumber\\
&& \left. \hspace{30 mm}
-4J_1^2(\kappa_q)+2J_2^2(\kappa_q)\right)\label{eq:anal}
\end{eqnarray}
where $E_0$ is the initial cloud energy, $\kappa_q=2\phi_d\sin(\kbar/2)$
and $J_\alpha(x)$ are the Bessel $J$ functions. The energies are given
in units of the recoil energy $E_R=\hbar^2k_l^2/(2m)$, and are periodic in
$\kbar$. This periodicity of the final energy is a feature that is not
reproduced in classical physics, and  
is a quantum phenomenon. For values of $\kbar$ which are multiples of
$2\pi$, a `resonance' appears, which becomes sharper for larger
numbers of kicks.

For sufficiently large kick strength $\phi_d$ and for values of $\kbar$ between
zero and the quantum resonance, or between the
quantum resonances, equations~\ref{eq:anal} describe a series of increases and
decreases in the final energy as a function of $\kbar$, forming {\em diffusion
resonances} \cite{andrew1}, which are due to quantum correlations in the atomic motion.
Note that these resonances are transient in nature, as
after a finite Heisenberg time, dynamic localisation will prohibit further energy growth and
that at larger kick numbers, these `resonances' will hence be less visible.
As the argument of the Bessel functions $\kappa_q$ depends on the intensity
of the laser light, any variation of the 
laser intensity over the atomic cloud will smear out these `resonances', 
also effectively lowering them. 

A large body of experimental \cite{Raizen99,exp} and theoretical
\cite{theo1,theo2,theo2b,theo3} work exists, which explores the dynamics of this system in
the limit of a large number of kicks. Generally, the energy growth
is measured for different kick periods and kick strengths, and compared
to numerical simulations.
The experiments start with an ensemble of atoms, typically ceasium or rubidium,
in magneto-optical trap at temperatures in the microKelvin range. This means that the
average momentum of the atoms before kicking is much larger than a single photon recoil. 
Starting from a single momentum state, the kicks distribute the atoms over many momentum states, spaced by two photon recoils. Interference between these momentum states eventually curb the growth of the momentum spread. It is therefore convenient to consider a quasimomentum $q=p\ {\rm mod}\ 2\hbar k$. At microKelvin temperatures, the atoms have a flat, even distribution of quasimomenta. The absolute momentum $p$, is of minor importance for the evolution of the energy spread. This means that the initial temperature of the MOT has very little effect on the results of the measurements. A detailed study of the effect of the initial momentum distribution can be found in reference \cite{mark2}.

In further experiments, the effects of amplitude and phase noise have been
studied \cite{ampnoise,phasenoise}, as well as the effect of spontaneous
emission
\cite{spontem,spontem2}. It was found that amplitude noise smoothed out the diffusion
resonances, while preserving the quantum resonance \cite{smoothdiffres}. Phase
noise in the kick
sequence was found to destroy the effects of dynamic localisation \cite{killqr}.
Spontaneous
emission was found to broaden the quantum resonance peak, making it easier
to observe experimentally for large numbers of kicks, which was discussed in a
recent paper by Wimberger {\em et.al.} \cite{Wimberger1}. They also mention a
remaining questions in the atom optics kicked rotor system, which concerns the
observation of higher order quantum resonances.

Here, we present the first experimental results 
showing the development of the diffusion resonances as a function of kick 
number  and for different kick strengths, and compare these directly with the 
analytical equations~\ref{eq:anal}. Although oscillations in the
mean energy have been observed before, they have only been studied
as a function of the kick strength \cite{Raizen98}. When the mean energy was
studied as function of the kick period, attention was focused on
the  quantum resonance \cite{darcy1}. Furthermore, we extend the range of
experimentally investigated kick periods to $\kbar=20\pi$ and demonstrate that
the final energy is periodic in $\kbar$ to that value.

In our experiment we trap rubidium 85 atoms in a standard magneto-optical 
trap~\cite{raab87}, loaded from a background vapour. After an 
accumulation phase, the anti-Helmholtz magnetic field is turned off and the trap 
lasers are switched to a larger detuning for a 3~ms cooling phase, after which 
the trap lasers are extinguished. The velocity distribution of the atoms at
that moment can be described by a Gaussian curve with a FWHM of
$\sim$4~cms$^{-1}$, which corresponds to a temperature of $\sim~10~\mu K$. The
atoms are then subjected to a periodic sequence 
of kicks by the, linearly polarised, standing wave laser field. The atomic 
ensemble is subsequently allowed to expand for 15~ms, after which the ensemble 
fluorescence is imaged on a CCD camera (Apogee AP47p) by flashing on the 
molasses lasers for 5~ms. We obtain the positional variances of the atomic 
cloud, in both the direction of the kick laser (`kicked') and orthogonal to that 
(`non-kicked'), numerically from the image and convert these to average velocity 
squared, and from that to kinetic energy. We take the energy in the non-kicked 
direction as the initial energy. The experiment is then repeated for different 
kick periods or kick numbers. The average initial energy is determined from an 
experimental run over a range of parameters, and is used to rescale the energy 
ratio to an energy. The kick laser beam was obtained from a Toptica DLX110 laser 
system, which was locked to the $D_2$ transition in $^{87}$Rb, obtaining a laser 
detuning of 1.3~GHz from the $F=3\rightarrow F=4$ transition in $^{85}$Rb. The 
beam was passed trough an AOM, which was controlled by a home-built programmable 
pulse generator, allowing us to pulse the laser with adjustable pulse number, 
period and amplitude. The pulse length for most experiments was set to 300~ns. 
To spatially filter the laser beam, it was then passed through a polarisation 
preserving single mode fibre. After collimation, the radius $(1/e^2)$ of the 
Gaussian laser beam was 3.0~mm. The initial radius of the atomic cloud was 
0.3~mm ($1\sigma$), small compared to the laser beam size. The laser beam was passed through a polarising 
beam splitter to ensure linear polarisation. The large detuning and linear laser 
polarisation yield equal kick strengths for all magnetic sublevels of the $F=3$ 
ground state. The maximum kick laser power was 100 mW, which for a 300~ns pulse 
and a 1.3~GHz laser detuning yields a scaled kick strength $\phi_d=5.2$. Larger 
kick strengths could be obtained by lengthening the kick pulse. 

The detuning of 
the kick laser yields a spontaneous emission probability of 1\% per atom per kick 
for this kick strength, which we deem negligible. This was verified by observing the 
momentum variance in the kicked direction for a range of 
kick numbers, up to 80 kicks. These experiments showed negligible energy growth 
after a certain number of kicks. This represents strong evidence that the quantum correlations,
which curb the energy growth, survive that number of kicks, justifying the neglect of spontaneous 
emission in our experiments. 

\begin{figure}
\begin{center}
\epsfig{file=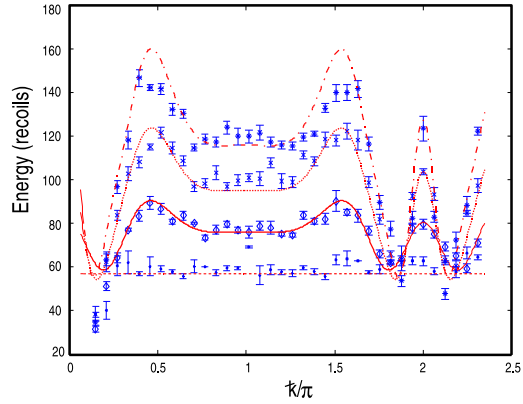, width=7cm}
\end{center}
\caption{The energy difference as a function
of $\kbar$, for $\phi_d=4.8$,  for two (dots), three (diamonds)
four (crosses) and five (stars) kicks. The lines represent the analytical
expressions, averaged over a 10\% intensity variation. \label{fig:rangekicks}}
\end{figure}

The first feature in eqns~\ref{eq:anal} that is of interest is the development 
of the structure in the final energy, from non-existent for two kicks, to quite 
pronounced for five kicks. In figure~\ref{fig:rangekicks} the energy difference 
between the `kicked' and `non-kicked' dimensions is shown as a function of the 
kick period for two to five kicks. For two kicks,the spectrum is flat as 
expected. This provides us with a reference value of $\phi_d$. For three kicks 
structure starts to appear, which is well reproduced by the analytic expressions 
in equation~\ref{eq:anal}. The quantum resonance at $\kbar=2\pi$ is starting to 
appear, and an onset of the `diffusion resonances' can be observed for smaller 
$\kbar$. For four and five kicks, these structures become more and more 
pronounced, but stay at the same position. The value of the kick strength 
$\phi_d=4.8$ for the analytical line is determined from the energy after two 
kicks. The analytical line takes a variation of 10\% in kick strength into 
account, which could be due to the Gaussian laser beam profile, or imperfect 
wavefronts in the kick laser beam. We also note the deviation of the first few 
points from the theoretical curve, as this is where the system exhibits more 
classical behaviour. This phenomenon is detailed in ref \cite{mark1}.

\begin{figure}
\epsfig{file=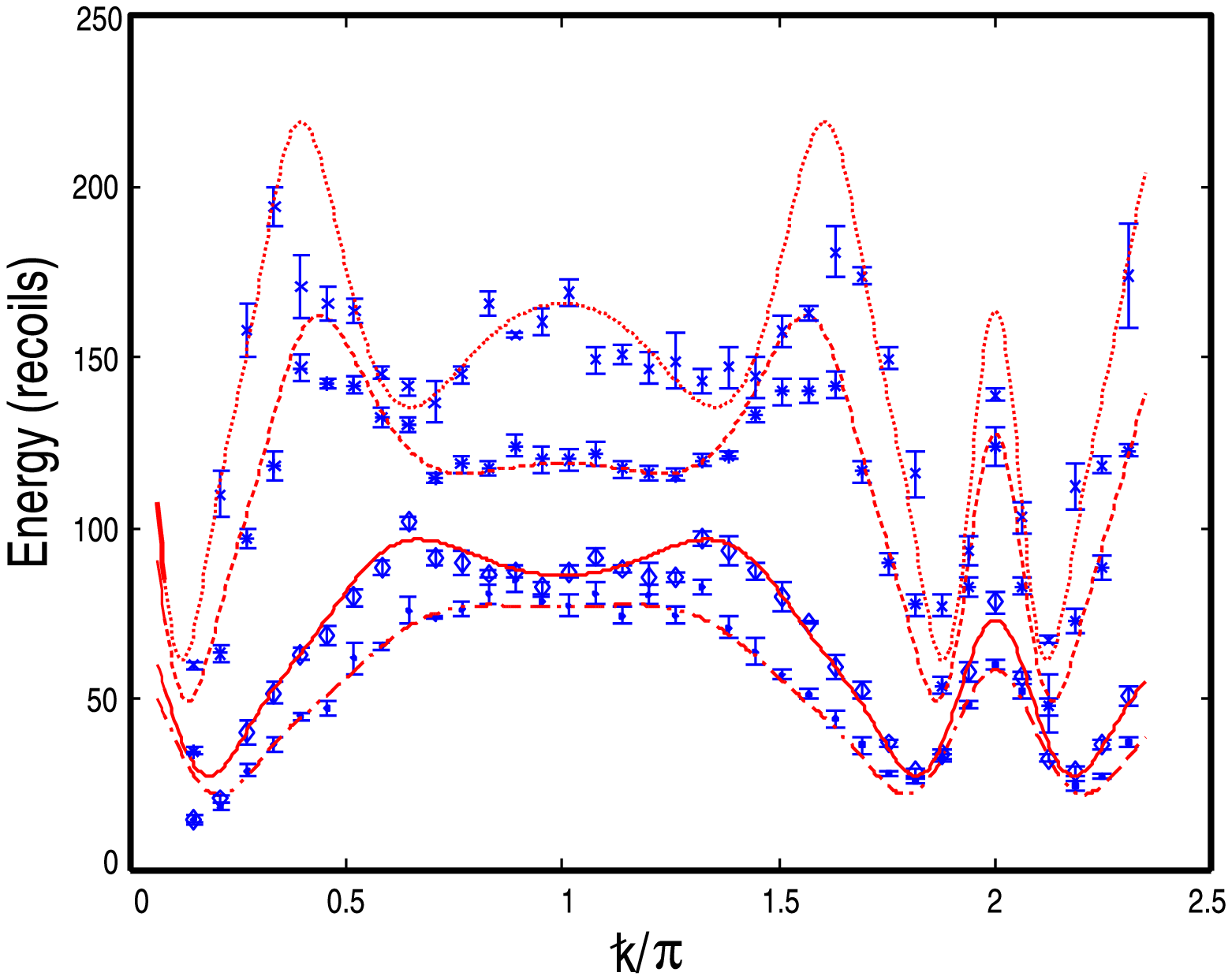, width=7cm}\epsfig{file=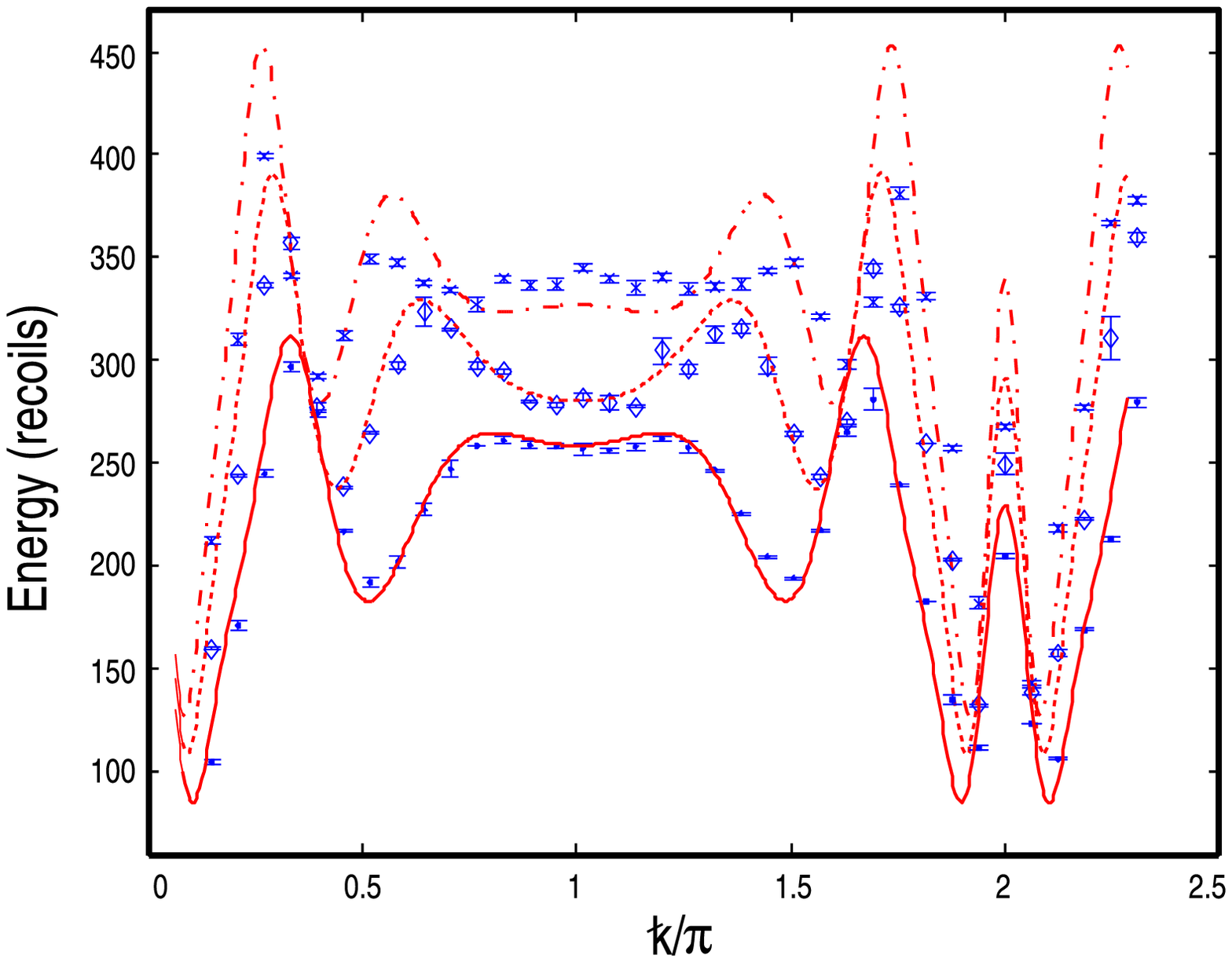, width=7cm}
\caption{The energy difference after five kicks as a function of $\kbar$ for
(a) lower kick strengths: $\phi_d=$3.4 (dots), 3.9 (diamonds), 5.0 (stars) and
5.7 (crosses) and (b) higher kick strengths: $\phi_d=$ 6.7 (dots), 7.6
(diamonds) and 8.2 (crosses). The lines
represent the analytical expression, averaged over a 10\% intensity variation.
\label{fig:phids}}
\end{figure}

As the argument of the Bessel functions $\kappa_q$ in eqns~\ref{eq:anal} is
proportional to the scaled kick strength $\phi_d$,
distinctly, we expect distinct qualitative differences in the final energy 
curves for different values of $\phi_d$.
In figure~\ref{fig:phids} the energy difference is shown as a function of
$\kbar$ for five kicks for a range of values for $\phi_d$. Again, the analytical
lines take a variation of 10\% in kick strength into account. For $\phi_d=3.4$,
virtually no diffusion resonances are observed. For larger values of $\phi_d$,
maxima are starting to appear, which move out at increasing $\phi_d$. An
additional maximum starts to form in the centre for $\phi_d=5.7$, in agreement
with the analytical formulae. At the same time, the quantum resonance gets more
narrow, as can be seen from the movement of the minima in the energy towards the
quantum resonance. As the energy at the minima remains close to the energy after
two kicks, which is proportional to $\phi_d^2$, this energy increases with
increasing $\phi_d$.

At even larger values of $\phi_d$, multiple oscillations of the final energy
as a function of $\kbar$ are expected. For the results in
figure~\ref{fig:phids}(b)
the pulses were lengthened to enable larger kick strengths. We clearly see 
multiple oscillations in the final energy as a function of $\kbar$ developing 
for these large kick strengths. Again, the lines represent eqns~\ref{eq:anal}, 
averaged over a 10\% spread in $\phi_d$. To our knowledge, this represents the 
first direct observation of multiple oscillations in the final energy, or
diffusion resonances.

\begin{figure}
\begin{center}
\epsfig{file=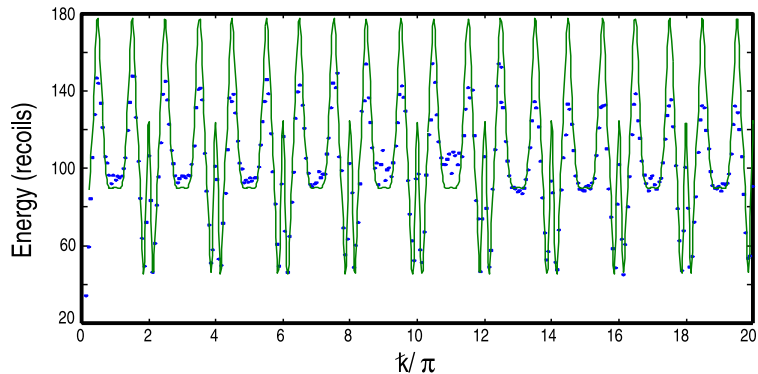, width=10cm} 
\end{center}
\caption{The energy difference after five kicks as a function of $\kbar$ for 
$\phi_d=5$, demonstrating the periodicity of the energy structure. The dots
represent experimental data, and the line represents the
analytical expression, taking a 10\% intensity fluctuation
into account.\label{periodic}}
\end{figure}

As indicated by Wimberger {\em et. al.}~\cite{Wimberger1}, an interesting
question concerns
the periodicity of the final energies that is 
apparent from equations~\ref{eq:anal}. This periodicity has been observed for
relatively large kick numbers and in the presence of spontaneous emission up to
a value for $\kbar=6\pi$ in reference~\cite{darcy1}. From a classical point of
view, long 
periods between the kicks should put the system in a chaotic regime, where kicks 
are completely unrelated. Quantum-mechanically, the induced coherence in the 
atomic motion should remain indefinitely, and there should be no difference in 
the physics between $\kbar$ and $\kbar+2M\pi$, where $M$ is a (positive) 
integer. In figure~\ref{periodic} we display the energy difference as a function 
of $\kbar$ for ten periods in $\kbar$, along with the analytical line, for 
$\phi_d=5$. The structure appears to be completely periodic, in sharp contrast 
to classical predictions, but in agreement with Shepelyansky's equations. Note 
that both the quantum resonance, at integer multiples of $2\pi$, which is a 
quantum phenomenon, and the diffusion resonances, which are thought to be 
classical in origin, appear to be completely periodic in $\kbar$. There is no 
apparent degradation of either the quantum- or diffusion resonances up to 
$\kbar=20\pi$. This demonstrates the quantum-mechanical nature of the system,
with no apparent decoherence.

In summary, we have experimentally verified the validity of Shepelyansky's 
expressions for the diffusion rate for the first five kicks in a quantum kicked 
rotor system. We find good agreement between experiment and the analytical 
expressions, we verify the periodicity of the results, and observe the predicted 
diffusion- and quantum resonances up to $\kbar=20\pi$.

\begin{acknowledgements}
The authors would like to acknowledge stimulating discussions
with S. Parkins, M. Sadgrove, T. Mullins, A. Hilliard and R. Leonhardt, as well 
as financial support from the University of Auckland.
\end{acknowledgements}

\end{document}